\documentclass[10pt,aps,twocolumn,prl,superscriptaddress,bibnotes,amsmath,amsfonts,floatfix,a4paper,longbibliography]{revtex4-2}

\usepackage{graphicx}
\usepackage{multirow}
\usepackage{verbatim}
\usepackage{amssymb,mathrsfs}
\usepackage{soul}
\usepackage{microtype}
\renewcommand{\thefigure}{\arabic{figure}}

\usepackage{lmodern}
\usepackage{textcomp}

\usepackage[pdfusetitle]{hyperref}

\newcommand{\I}{\mathrm{i}}

\usepackage[dvipsnames]{xcolor}

\begin{document} 

\title{2D coherent spectroscopy signatures of exciton condensation in Ta$_2$NiSe$_5$}
\author{Jiyu Chen}
\affiliation{Institute of Physics, Chinese Academy of Sciences, Beijing 100190, China}
\affiliation{Songshan Lake Materials Laboratory, Dongguan, Guangdong 523808, China}
\affiliation{Department of Physics, University of Fribourg, 1700 Fribourg, Switzerland}
\author{Jernej Mravlje}
\affiliation{Jožef Stefan Institute, Jamova 39, SI-1000 Ljubljana, Slovenia}
\affiliation{Faculty of Mathematics and Physics, University of Ljubljana, Jadranska 19, 1000 Ljubljana, Slovenia} 
\author{Denis Gole\v z}
\affiliation{Jožef Stefan Institute, Jamova 39, SI-1000 Ljubljana, Slovenia}
\affiliation{Faculty of Mathematics and Physics, University of Ljubljana, Jadranska 19, 1000 Ljubljana, Slovenia} 
\author{Philipp Werner}
\affiliation{Department of Physics, University of Fribourg, 1700 Fribourg, Switzerland}

\date{\today}

\begin{abstract}
We show that the nonlinear optical response probed by two-dimensional coherent spectroscopy (2DCS) can discriminate between excitonic and lattice driven order. In the excitonic regime of a realistic model of Ta$_2$NiSe$_5$, the third order 2DCS signals are strongly enhanced by the condensate's amplitude and phase modes, with negligible contributions from single-particle excitations. In the linear optical response, in contrast, single-particle and collective-mode contributions overlap. With increasing electron-phonon coupling, the amplitude mode contribution to 2DCS  initially remains robust, but then drops rapidly and remains small in the phonon-dominated regime -- even in systems with large order parameter. 2DCS also aids the detection of the massive relative phase mode, which is analogous to the Leggett mode in superconductors. 
Our analysis, based on the time-dependent Hartree-Fock approach, demonstrates that 2DCS can track the emergence of the symmetry-broken state and the crossover from Coulomb-driven to phonon-driven order. 
\end{abstract}

\maketitle
{\it Introduction -- }
An exciton is a charge neutral object formed by an electron-hole pair bound by the Coulomb interaction. 
Excitons can spontaneously condense in the presence of strong electron-hole attraction~\cite{cloizeaux1965,jerome1967,kunes2015}, leading to an excitonic insulator (EI). In artificial semiconducting quantum wells~\cite{eisenstein2014,eisenstein-MacDonald2004,Nandi2012} or bilayer heterostructures~\cite{paik_excitons_2024,nguyen2025,qi2025,wang2019,ma2021}, the electrons and holes may be located in different layers, and hence  separated by insulating barriers. Since the dynamics of the excitons is distinct from that of individual electrons and holes, 
an exciton condensation leads to exotic responses, such as interference patterns in photoluminescence~\cite{butov_macroscopically_2002} or perfect Coulomb drag~\cite{Nandi2012,nguyen2025,qi2025}.
However, most naturally occurring excitonic insulator (EI) candidates are semimetals or narrow-band semiconductors, such as 1$T$-TiSe$_2$~\cite{cercellier2007,monney2010,kogar2017}, Ta$_2$Pd$_3$Te$_5$~\cite{zhang2024,hossain2025,huang2024,yao2024}, and 
Ta$_2$NiSe$_5$ (TNS)~\cite{wakisaka2009,kaneko2012,windgatter2021}. 
Since bulk materials do not allow the use of ideal probes of excitonic condensation—such as perfect Coulomb drag—it is crucial to identify experimental techniques that can provide unambiguous evidence for excitonic condensation.

The widely studied EI candidate TNS exhibits a phase transition from a high-temperature semi-metal to a low-temperature small-gap ($\sim0.16$~eV) semiconductor~\cite{disalvo1986,lu2017,larkin2017,larkin2018} at $T_{c}=328$~K. It is accompanied by a structural phase transition from the high-temperature orthorhombic phase to a low-temperature monoclinic phase, which breaks the reflection symmetries~\cite{mazza2020,watson2020,chatterjee2025}. On the one hand, the flattened top of the valence band~\cite{wakisaka2009,wakisaka2012,seki2014} seen in photoemission spectroscopy, peculiar fluence dependencies observed in pump-probe spectroscopy~\cite{mor2017,mor2018,okazaki2018,bretscher2021,saha2021}, softening of the electronic excitations in Raman spectroscopy~\cite{volkov2021,volkov2021failed}, and the almost vanishing thermopower at low temperature~\cite{nakano2019} point to excitonic condensation. On the other hand, the broadening~\cite{kim2021,ye2021} and softening~\cite{kim2020} of the B$_{2g}$ phonon mode reported by Raman spectroscopy also indicates a key role of the phonon coupling~\cite{watson2020,subedi2020,windgatter2021}. Recent theoretical studies suggest that the low-temperature state features a mixture of excitonic and lattice order parameters and highlight the importance of 
clarifying the interplay 
between the two~\cite{mazza2020,kaneko2012,chatterjee2025}.

Following an approach similar to Higgs spectroscopy in superconductors~\cite{matsunaga2013,matsunaga2014,katsumi2017,grasset2019,shimano2020,schwarz2020}, one strategy to quantify the competition between different degrees of freedom is to probe collective excitations—such as the amplitude mode or the massive phase mode—which act as sensitive indicators of such competition~\cite{murakami2020}. Experimental studies so far have employed Raman spectroscopy~\cite{kim2021,ye2021,volkov2021,volkov2021failed,katsumi2023}, nonlinear transport measurements~\cite{bretscher2021b}, and optical spectroscopy. On the theoretical side, nonlinear optical probes~\cite{golez2020,osterkorn2024} and high-harmonic generation~\cite{tanabe2021} have been proposed. A major challenge, however, lies in distinguishing these collective modes from single-particle excitations, given their often overlapping energy scales.

In this Letter, we theoretically investigate how exciton condensation in TNS manifests itself in two-dimensional coherent spectroscopy (2DCS)~\cite{mukamel1995,hamm2011} measurements. 2DCS is a technique which has been recently introduced to study coherent spin dynamics~\cite{wan2019,li2021,li2023,gao2023,li2024}, inhomogeneity~\cite{liu2024}, nonequilibrium states~\cite{chen2025} and collective modes~\cite{chen2025b} in quantum materials~\cite{liu2025}.
It employs a sequence of weak, phase-stable broadband pump pulses to extract nonlinear information on the excitation-deexcitation pathways. Both theoretical~\cite{chen2025b,tsuji2025} and experimental~\cite{katsumi2024,katsumi2025} works demonstrated that 2DCS can identify the amplitude mode in weak-coupling antiferromagnets and superconductors.

Our calculations show that the nonlinear 2DCS signal is well suited for detecting  exciton(-lattice)
condensation in TNS. In particular, 2DCS is sensitive to the mechanism which drives the symmetry breaking, and hence allows to track the crossover from exciton-dominated to phonon-dominated order. We furthermore reveal the signatures of a relative phase mode analogous to the Leggett mode in multiband superconductors~\cite{leggett1966}.

\begin{figure}[t]
	\includegraphics[width=1.0\linewidth]{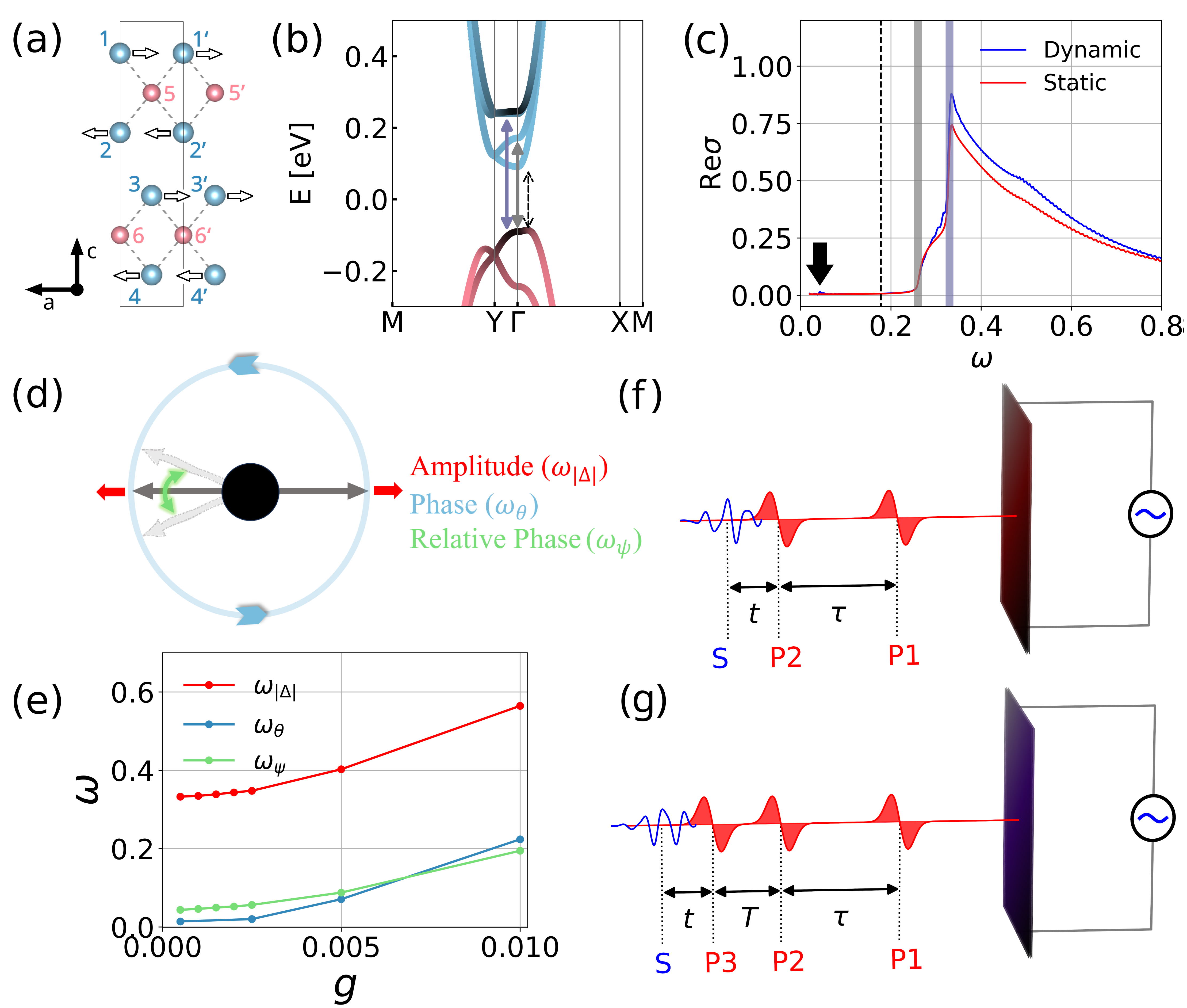}
	\caption{
		(a)~Unit cell of TNS with Ta~(blue) and Ni~(red) atoms. Black arrows indicate the B$_{2g}$ distortion. 
		(b)~Band-structure in the EI phase. The blue~(red) shading encodes the Ta~(Ni) character of the bands.
		(c)~Optical conductivity Re$\sigma$ for a weak laser excitation to the ordered state, obtained with a dynamic and static Fock term (and damping factor $\eta=0.003$~eV in the Fourier transform). 
		The vertical dashed black, solid gray and solid violet lines indicate the gap energies marked by the corresponding arrows in (b).
		The black arrow indicates the relative phase mode.
		The results in (b,c) are for $V=0.784$~eV, $g=0.0005$~eV and $\beta=100$~eV$^{-1}$. 
		(d) Illustration of the amplitude, phase and relative phase mode, with the black arrows representing Ni-Ta hoppings. (e) Evolution of the mode energies with $g$ for fixed single-particle gap ($\beta=100$~eV$^{-1}$, $V=0.784$~eV for $g\le 0.0025$~eV, $V=0.781$~eV for $g=0.005$~eV, $0.7725$~eV for $g=0.01$~eV). 
		(f,g) Pump excitations (red) in the two-pulse (f) and three-pulse (g) setup combined with the optical current measurement (blue line). 
        } \label{fig1}
\end{figure}

{\it Model and method -- } To study the layered material Ta$_2$NiSe$_5$ with quasi-one-dimensional Ta-Ni-Ta chains, we use a six band low-energy model containing 4 Ta orbitals and 2 Ni orbitals per unit cell with lattice vectors $a=3.503$~\AA~and $c=15.761$~\AA, as labeled in Fig.~\ref{fig1}(a). Following Refs.~\cite{mazza2020,chatterjee2025}, the spinor operator $\hat \Psi_{\mathbf{R}\sigma}=\{\hat c_{1\sigma \mathbf{R}},\ldots,\hat c_{6\sigma \mathbf{R}}\}$ is introduced to define the tight-binding Hamiltonian
\begin{equation}
	\label{eq:tb}
\hat H_{\text{kin}}=\sum_{\mathbf{R}\boldsymbol{\delta}} \hat \Psi^{\dagger}_{\mathbf{R}+\boldsymbol{\delta}\sigma} h(\boldsymbol{\delta},t) \hat \Psi_{\mathbf{R} \sigma}, 
\end{equation}
where $\hat c_{i\sigma \mathbf{R}}$ is the annihilation operator of orbital $i$ and spin $\sigma$ in the unit cell indexed by $\mathbf{R}$. The hopping matrix $h(\boldsymbol{\delta},t)$ connects it with the neighboring cell indexed by $\mathbf{R}+\boldsymbol{\delta}$ and the time-dependence is given by the Peierls substitution, as explained below.

In the interaction term, we consider the on-site Hubbard interaction $U$ for all the Ta and Ni atoms, and the nearest neighbor interaction $V$ between the Ta and Ni atoms,
\begin{align}
	\hat H_{\text{int}}(\mathbf{R})=&\,\, U \sum_{1\leq i \leq 6} \hat n_{i\uparrow,\mathbf{R} }\hat n_{i\downarrow,\mathbf{R}} +  V \sum_{\substack{ij=15,25,36,46}}  \hat n_{i\mathbf{R}} \hat n_{j \mathbf{R}}\nonumber\\
	&+V \sum_{i=1,2}  \hat n_{i\mathbf{R+a}} \hat n_{5 \mathbf{R}}+V \sum_{i=3,4}  \hat n_{i\mathbf{R-a} }\hat n_{6 \mathbf{R}},
\end{align}
where $\hat{n}_{i\mathbf{R}}=\sum_{\sigma=\uparrow,\downarrow}{\hat{n}_{i\sigma\mathbf{R}}}=\sum_{\sigma=\uparrow,\downarrow}{\hat{c}^\dagger_{i\sigma\mathbf{R}}\hat{c}_{i\sigma\mathbf{R}}}$ defines the charge density of orbital $i$ in unit cell $\mathbf{R}$.
To be consistent with the monoclinic distortion, we break the reflection symmetries $\sigma_{\parallel,\perp}^{A/B}$ by introducing a coupling to Einstein phonons $\hat H_{\text{ph}}(\mathbf{R})=\frac{\omega_0}{2} (\hat X^2_\mathbf{R}+\hat \Pi^2_\mathbf{R})$ with frequency $\omega_0$. Here, $\hat X_\mathbf{R}~(\hat \Pi_\mathbf{R})$ is the displacement~(momentum) operator of the Holstein phonon, which couples to the six orbitals of the unit cell $\mathbf{R}$  through the $B_{2g} $ channel~ \cite{mazza2020,chatterjee2025}
\begin{equation}\label{Eq:elph}
	\hat H_{\text{el-ph}}(\mathbf{R})=g \hat X_\mathbf{R} \hat\Delta_\mathbf{R} .
\end{equation}
The operator $\hat\Delta_\mathbf{R} \equiv \hat \phi_{15\mathbf{R}}-\hat \phi_{25\mathbf{R}}-\hat \phi_{36\mathbf{R}}+\hat \phi_{46\mathbf{R}}$ defines the $B_{2g}$ order parameter $\Delta_\mathbf{R}=\langle \hat \Delta_\mathbf{R}\rangle\equiv \Delta$, which minimizes the energy of the system at equilibrium~\cite{chatterjee2025b}. For the orbital pair $\{ij\}=15,25~(36,46)$,
$\hat \phi_{ij\mathbf{R}}= \sum_{\sigma}\hat c_{i\sigma\mathbf{R}}^{\dagger} \hat c_{j\sigma\mathbf{R}}+\hat c_{i\sigma\mathbf{R\pm a}}^{\dagger} \hat c_{j\sigma\mathbf{R} }$, with sign $+$ for 15,25 (sign $-$ for 36,46) in the second term.
We use time-dependent Hartree-Fock theory to simulate the evolution of the system at the mean-field level (see SM) and compare the response with 
and without the self-consistent evaluation of the Hartree-Fock term. The former ``dynamic" treatment includes vertex corrections in the excitonic channel and the effect of collective modes, while the latter ``static" calculation  is equivalent in the weak-field limit to the bare bubble response in the excitonic channel, see Refs.~\cite{murakami2020,murakami2017,golez2020}.

We study the optical current induced by a sequence of single-cycle~(broadband) laser pulses, as  illustrated in Fig.~\ref{fig1}(f,g). Specifically, we consider setups with two (Fig.~\ref{fig1}(f)) and three (Fig.~\ref{fig1}(g)) collinear pump pulses. The time delay between the first and second pump pulse is $\tau$, and in the case of three pump pulses, the time delay between the second and third pump pulse is $T$. The induced current is measured in the time domain, with a delay $t$ relative to the last pump pulse. We scan over the time delay $\tau$ and subtract the contributions from single pump pulses (and two pump pulses in the three-pump case). Fourier transformation of the nonlinear current from the $(\tau,t)$ domain to the frequency domain yields the two dimensional spectrum in the $(\omega_\tau,\omega_t)$ space (for a fixed waiting time $T$ in the three-pulse setup).  
 In our real-time simulations, we use the Peierls substitution to incorporate the electric field $\mathbf{E}$ of the laser~\cite{peierls1933}. In the presence of a time-dependent electric field  $\mathbf{E}(t)$, the hopping amplitude connecting the two orbitals centered at $\mathbf{r}_m$ and $\mathbf{r}_m$ is replaced by $h_{mn}\to h_{mn}e^{i\phi_{mn}(t)}$. 
 The exponential factor $\phi_{mn}(t) = \int_0^t d t'\mathbf{E}(t')\cdot (\mathbf{r}_m-\mathbf{r}_n)$ is the Peierls phase.
The pulse $\mathbf{E}(t)=E_0\hat {\mathbf{e}}_a\frac{t-t_0}{\sqrt{2\pi}\sigma^2} e^{-\frac{(t-t_0)^2}{2\sigma^2}} $ with $E_0=300~$kV/cm has width $\sigma=1.3$~fs and is polarized along the Ta-Ni-Ta chains ($\hat{\mathbf{e}}_a$). 
 Its central frequency is $\omega = 1/\sigma =0.5$~eV. Below, we use the natural unit system with $k_B$ = $\hbar$ = 1. When not explicitly stated, the unit of energy is eV. 

{\it Results --}
\label{sec:result}
We set the onsite interaction to $U=2.5$ eV and the phonon frequency to $\omega_0=8$~meV (2~THz), as suggested by Refs.~\onlinecite{mazza2020,chatterjee2025}, and focus on systems in or near the ordered state. 
Figure~\ref{fig1}(b) depicts the band structure for $V=0.784$ eV, inverse temperature $\beta=100$ eV$^{-1}$, and a weak electron-phonon coupling $g=0.0005$~eV. The orbital character of the bands is indicated with colors. 
The excitonic hybridization between the Ta atoms and the valence band is strongest for the high-lying conduction band and near the $\Gamma$ point.

The (linear) optical conductivity is shown by the blue line in Fig.~\ref{fig1}(c).
To clarify the effect of the feedback from the dynamics of the excitonic condensate 
we also calculate the time evolution with the Fock term fixed to its initial value without electric field~\cite{tsuji2020,chen2025b} (red line in Fig.~\ref{fig1}(c)). With the static Fock protocol, the oscillations of the excitonic order parameter are strongly suppressed compared to the calculation with dynamic Fock term (see SM Fig.~S1), so the
 difference between the blue and red optical conductivity spectra illustrates the effect of the collective excitonic response. In both spectra, the optical gap of 0.26 eV is controlled by the direct gap between the top valence band and the middle conduction band. (Transitions to the lowest conduction band must be suppressed by symmetry.) Furthermore, the conductivity exhibits a sharp resonance at $\omega\approx 0.33$ eV $\equiv \omega_{|\Delta|}$, corresponding to the gap between the strongly hybridized bands, and a shoulder near $\omega=0.5$~eV 
matching the transitions near $\Gamma$ between the lowest Ni and highest Ta bands. 
The enhancement of the resonance by collective modes is clearly visible but in experiments difficult to disentangle from the dominant contribution associated with single-particle excitations~\cite{larkin2017}. 
The small peak at $\omega\approx 0.04$~eV~$\equiv \omega_\psi$ in the response with vertex corrections (see black arrow) originates from a massive
{\it relative} phase mode (Fig.~\ref{fig1}(d)), analogous to the Leggett mode in multi-band superconductors \cite{leggett1966}, as discussed in the Supplemental Material (SM). 
Its energy depends on the strength of the hybridization between the Ta and Ni orbitals and the electron-phonon coupling $g$~\cite{chatterjee2025,sun2021} (Fig.~\ref{fig1}(e)).

\begin{figure}[t]
	\includegraphics[width=1.0\linewidth]{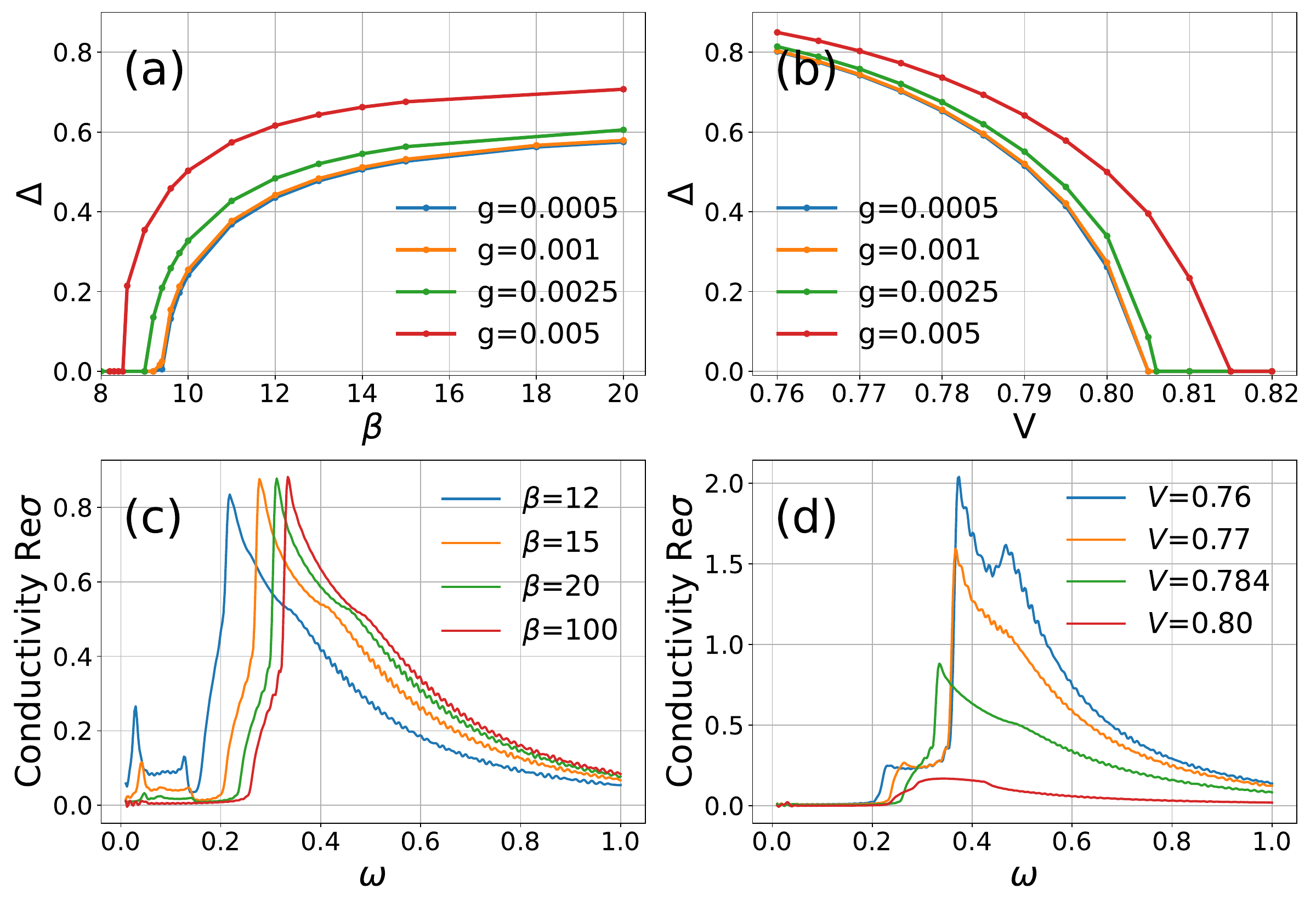}
	\caption{(a) Order parameter $\Delta$ as a function of inverse temperature for $V=0.784$ eV 
	and (b) as a function of the inter-site interaction $V$ for $\beta = 100$ eV$^{-1}$ and various el-ph couplings $g$. (c) Conductivity as a function of inverse temperature $\beta$ for $V=0.784$ eV 
	and (d) as a function of $V$ for $\beta = 100$ eV$^{-1}$ ($g=0.0005$~eV). A damping factor $\eta = 0.003$~eV is used in the Fourier transforms. See SM Fig.~S4 for the $g=0.0025$~eV results and the permittivity as a function of $\beta$.
	}
	\label{fig2}
\end{figure}

The excitonic order is sensitive to temperature and changes in the interaction parameters. As shown in Ref.~\cite{matsunaga2014}, it exists only in a narrow range of inter-site interactions $V$. 
Figure~\ref{fig2}(a,b) plots the temperature and $V$ dependence of the order parameter $\Delta$ in our setup. 
The optical conductivity reveals a gradual decrease of the optical gap with increasing temperature, and a clear shift and decrease of the main resonant peak, see Fig.~\ref{fig2}(c). At the same time, we observe an increase in the weight of the low-energy (relative phase mode and phase mode) signals, 
in qualitative agreement with experiment \cite{larkin2017}.
Increasing the nonlocal interaction $V$ reduces the excitonic order (Fig.~\ref{fig2}(b)) and suppresses the resonant peak in the linear optical conductivity (Fig.~\ref{fig2}(d)).

\begin{figure*}[t]
	\includegraphics[width=1.0\linewidth]{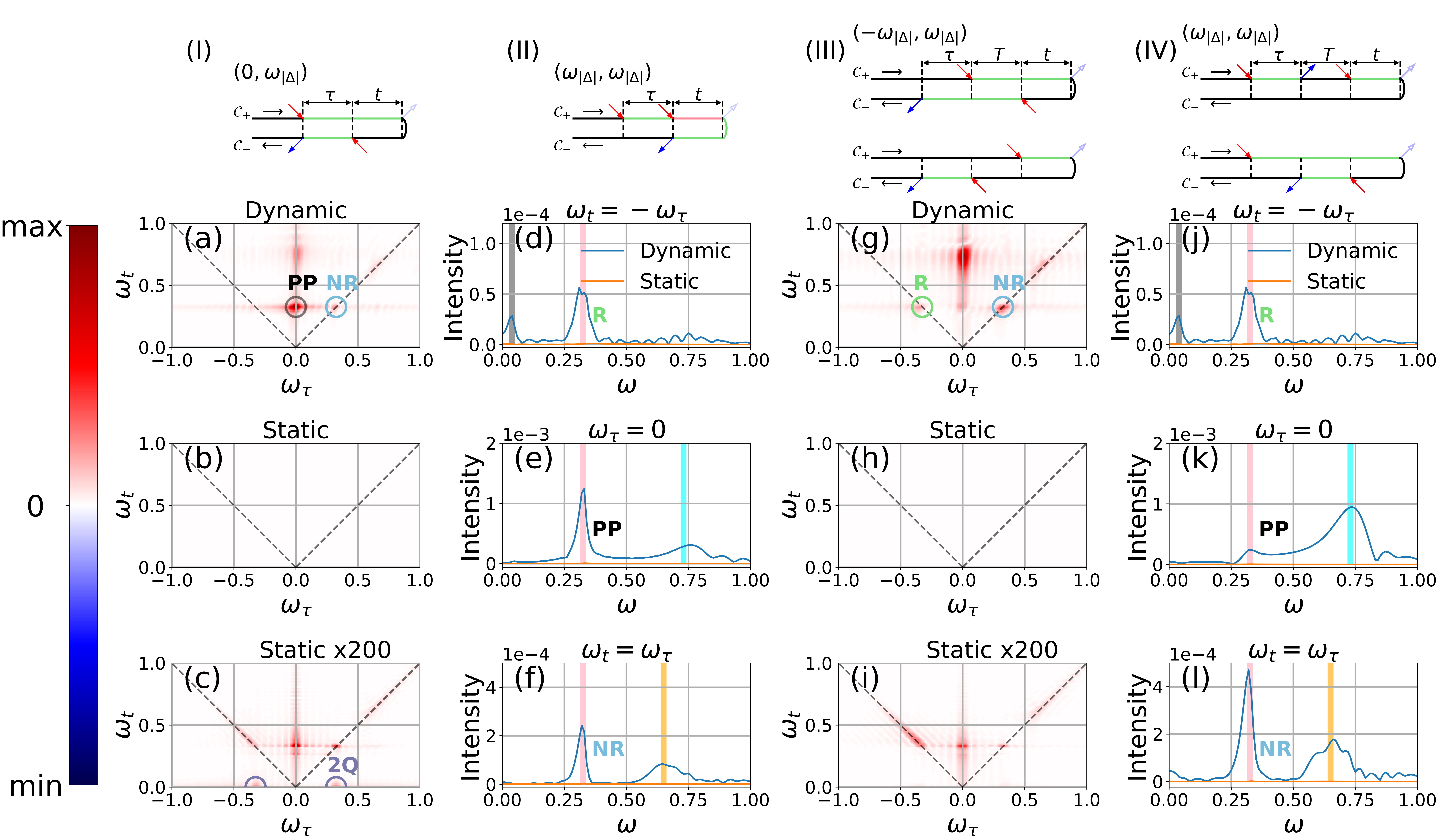}
	\caption{ 2DCS signals obtained using the two-pulse (a-c) and three-pulse (g-i) setups. (a,g) Results of the dynamic simulation with pump-probe~(PP), non-rephasing~(NR) and rephasing~(R) contributions highlighted. (b,h) Analogous spectra for the static simulations (with suppressed amplitude mode contribution), and (c,i) with intensities rescaled by a factor of 200. (d,j) Cuts of the 2DCS map along the antidiagonal ($-\omega_\tau=\omega_t$), (e,k) along the y-axis ($\omega_\tau=0$) 
	and (f,l) along the diagonal ($\omega_\tau=\omega_t$). The Keldysh diagrams in panels (I)-(IV) illustrate possible light-matter interaction pathways corresponding to the signals at the indicated $(\omega_\tau,\omega_t)$, see text for details. Vertical lines indicate different combinations of $\omega_{|\Delta|}$ and $\omega_{\psi}$ (see text). Parameters: $\beta=100$~eV$^{-1}$, $g=0.0005$~eV and $V=0.784$~eV. In the three-pulse measurement, the waiting time is $T=6.6$~fs.
	}
	\label{fig3}
\end{figure*}

In Fig.~\ref{fig3}, we show 2DCS spectra calculated with dynamic (a,g) and static (b,c,h,i) Fock terms. 
The signal intensity from the static calculation is suppressed by more than two orders or magnitude, which shows that the 2DCS signal of the excitonic insulator is completely dominated by the amplitude mode of the order parameter. 
We also show three cuts of these 2D spectra: along the anti-diagonal ($-\omega_\tau=\omega_t$) in panels (d,j), along the y-axis ($\omega_\tau=0$) in panels (e,k) and along the diagonal ($\omega_\tau=\omega_t$) in panels (f,l). 
The $0.33$~eV energy associated with prominent structures (pink vertical lines) corresponds to the energy $\omega_{|\Delta|}$ of the sharp resonance in the linear conductivity. In particular, this amplitude mode
energy defines the locations of the dominant rephasing (R) signal at $(\omega_\tau,\omega_t)=(-\omega_{|\Delta|},\omega_{|\Delta|})$ and nonrephasing (NR) signal at  $(\omega_\tau,\omega_t)=(\omega_{|\Delta|},\omega_{|\Delta|})$. The subdominant NR signal is at $(\omega_\tau,\omega_t)=(2\omega_{|\Delta|},2\omega_{|\Delta|})$ (orange vertical line in panel (f)). 
Panels (d,j) furthermore demonstrate that the R signal exhibits a low-energy peak at $(\omega_\tau,\omega_t)=(-\omega_\psi,\omega_\psi)$ (grey vertical lines), i.e., at the energy of the relative phase mode $\omega_\psi=0.04$~eV. 
In panel (e), the main pump-probe (PP) signal obtained with the two-pulse setup  is a third order response at $(\omega_\tau,\omega_t)=(0,\omega_{|\Delta|})$  (pink vertical line), while the broad contribution near $(\omega_\tau,\omega_t)=(0,2\omega_{|\Delta|}+2\omega_\psi)$ (cyan vertical lines) is associated with higher order processes. 
In the three-pulse setup (panel (k)), these higher order contributions dominate the PP signal.

The 2DCS spectra from the static calculation, magnified by a factor of 200 in panels (c,i), exhibit a somewhat different structure. In particular, we observe peaks at $(\omega_\tau,\omega_t)=(0,0.26~\text{eV})$, corresponding to the energy of the gap in Re$\sigma$ (gray line in Fig.~\ref{fig1}), a broad band associated with intra-band relaxation near R, and so-called two-quantum (2Q) signals near $(\omega_\tau,\omega_t)=(\pm\omega_{|\Delta|},0)$. These features are generally expected for a correlated insulator \cite{chen2025,chen2025b}, but in the dynamical calculation, they are completely overwhelmed by the sharp R, NR and PP signals originating from the collective modes.

These results demonstrate that the 2DCS signal can be used to detect spontaneous symmetry breaking into a (phonon-assisted) excitonic condensate.
Furthermore, with different phase matching processes (R/NR/PP) and excitation setups (two-pulse or three-pulse), one can extract the energies of the amplitude and relative phase modes and gain insights into higher-order nonlinear responses of the condensate.

It is instructive to draw Keldysh diagrams~\cite{chen2025}, which are equivalent to the double-sided Feynman diagrams in the quantum chemistry literature~\cite{mukamel1995}, to understand the sequence of underlying light-matter interactions. As shown in Fig.~\ref{fig3}(I), a possible two-pulse sequence for the PP signal is that the first pulse excites the amplitude mode (from black to green, excitation energy $\omega_{|\Delta|}$) on both branches of the Keldysh contour, while the second pulse deexcites it on one branch. 
Note that in contrast to the Higgs mode in superconductors~\cite{shimano2020,matsunaga2014}, the amplitude mode of the EI can be excited with a single photon process~\cite{murakami2020,golez2020}. During the first interval $\tau$, the system remains in a population state (without phase accumulation) before the second laser creates a superposition state (with phase accumulation). Hence, a total phase of $e^{-i\omega_{|\Delta|} t}$ is accumulated before the current measurement at time $t$ after the second laser pulse, and upon Fourier transformation from $(\tau,t)$ to $(\omega_\tau,\omega_t)$, we obtain the PP signal at $(\omega_\tau,
\omega_t)=(0,\omega_{|\Delta|})$ in the 2D spectrum. 
In contrast, in the light-matter interaction sequence illustrated in Fig.~\ref{fig3}(II), the first laser only excites the upper branch with energy $\omega_{|\Delta|}$, and thus during the interval $\tau$, a phase $e^{-i\omega_{|\Delta|} \tau}$ is accumulated. The second laser interacts with both branches, creating a double excitation in the upper branch (from green to pink) and exciting the lower branch (from black to green). Therefore, another phase $e^{-i\omega_{|\Delta|} t}$ is accumulated before the measurement takes place at time $t$ after the second pulse. This diagram corresponds to the NR signal at $(\omega_\tau,
\omega_t)=(\omega_{|\Delta|},\omega_{|\Delta|})$. For the three-pulse setup with intervals $\tau$, $T$ and $t$ between the three pump pulses and the current measurement, we plot two diagrams for the R and two diagrams for the NR signals in Fig.~\ref{fig3}(III) and \ref{fig3}(IV), respectively.  

\begin{figure}[t]
	\includegraphics[width=1.0\linewidth]{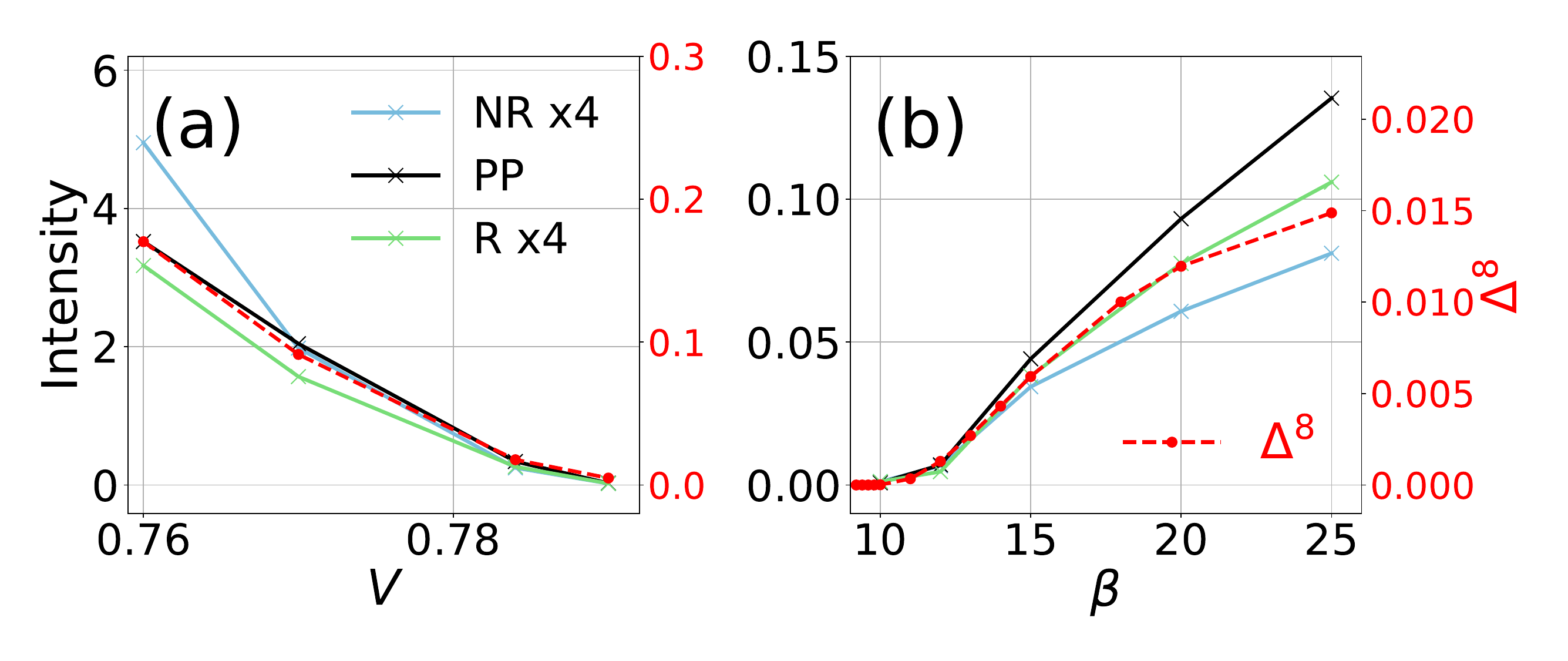}
	\caption{Intensity of the 2DCS signals (left axis) and 8th power of the order parameter $\Delta$ (right axis). NR signals are measured at $(\omega_\tau,\omega_t)=(\omega_{|\Delta|},\omega_{|\Delta|})$ and PP signals are measured at $(0,\omega_{|\Delta|})$ using the two-pulse setup. R signals are measured at $(-\omega_{|\Delta|},\omega_{|\Delta|})$ using the three-pulse setup. (a) Signal intensity for different $V$ with $g=0$, $\beta = 100$ eV$^{-1}$.  (b) Signal intensity for different $\beta$ with $g=0.0005$~eV, $V=0.784$~eV.}
	\label{fig4}
\end{figure}

Since the NR, R and PP signals in 2DCS have a dominant contribution from the collective mode of the EI, they should be useful to detect the presence of a condensate. 
Figure~\ref{fig4} shows the temperature and $V$ dependence of the three amplitude mode related peaks.
There is a clear correlation between the order parameter $\Delta$ and the intensities of the signals, which are obtained by integrating the spectra in a frequency window covering the peaks.
The numerical data suggest that near the critical temperature, the peak intensities scale like the 8th power of $\Delta$.

\begin{figure}[t]
	\includegraphics[width=\linewidth]{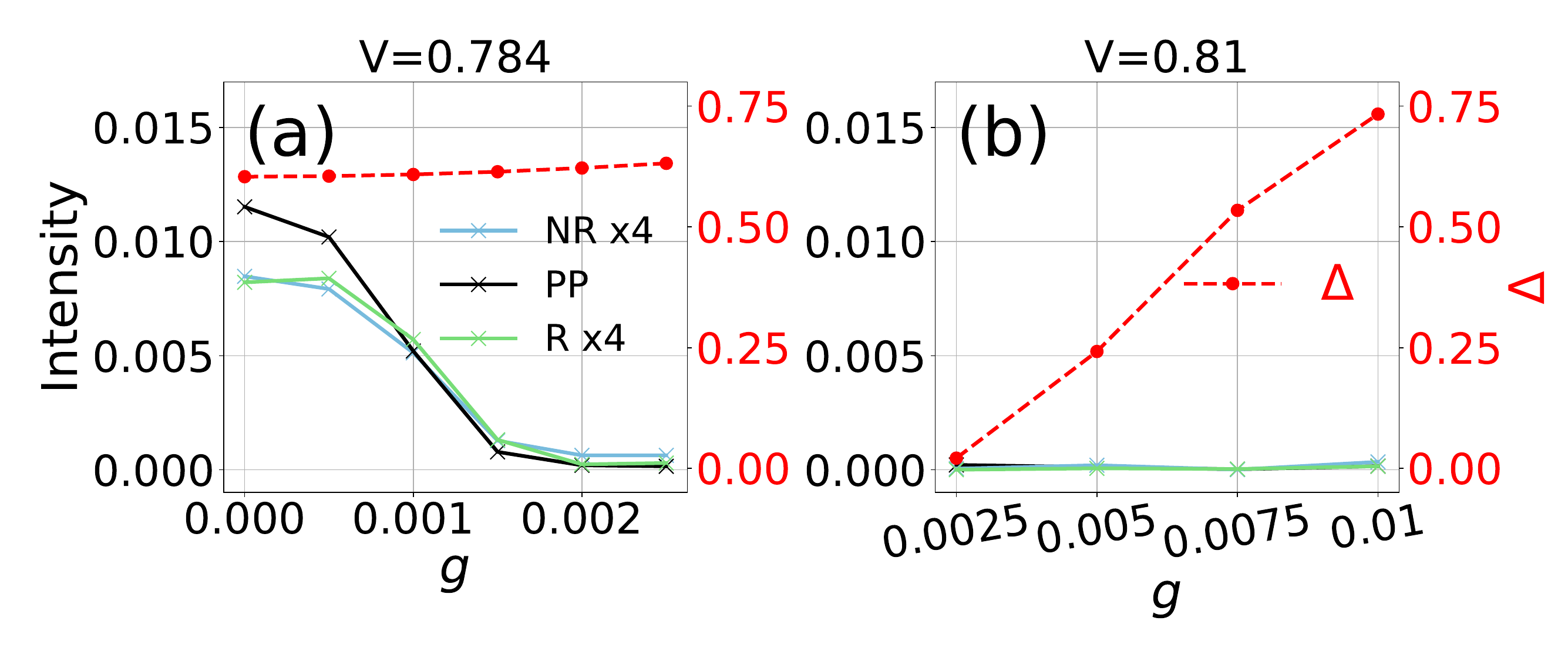}
\caption{Dependence of the amplitude mode dominated 2DCS signals (left axis) and of the order parameter $\Delta$ (right axis) on the phonon coupling $g$ for $\beta=100$ eV$^{-1}$ and for $V=0.784$~eV (a) and $V=0.81$~eV (b). The NR and PP intensities are obtained using the two-pulse protocol, while the R intensity is obtained using the three-pulse protocol.
}
	\label{fig5}
\end{figure}

In Fig.~\ref{fig5}(a), we plot the weight of the amplitude mode dominated R, NR and PP signals and the order parameter $\Delta$ as a function of the phonon coupling strength $g$ for $V=0.784$ eV and $\beta=100$ eV$^{-1}$. This figure illustrates the crossover from exciton-dominated to phonon-dominated order. At $g=0$, the system is a pure EI, and the large intensity of the 2DCS signals originates from the amplitude mode of this phase. Once the phonon coupling increases beyond $g\simeq 0.001$~eV, the weight of the 2DCS signals drops rapidly, 
suggestive of a change in the main driving force behind the order. At the same time, the growing phonon distortion leads to a mild increase in the order parameter, consistent with previous studies~\cite{chatterjee2025}. 
In Fig.~\ref{fig5}(b), we show analogous results for $V=0.81$~eV, where without electron-lattice interaction the system would be a normal semiconductor and the order parameter vanishes for $g=0.0005$~eV (Fig.~\ref{fig2}(b)). For much larger el-ph couplings, an order parameter with similar amplitude to the previous setup ($\Delta\sim 0.6$) is induced. In this phonon-dominated regime, however, the R, NR and PP signal intensities remain very weak, compared to the system with Coulomb-driven order (panel (a)).
In the generic case, the spontaneous electronic ordering and phonon distortion act cooperatively, but our results show that the 2DCS signal is selectively sensitive to exciton-dominated order.

\textit{Summary --} 
We demonstrated the power of 2DCS as a collective mode spectroscopy by simulating multi-pulse excitations of the EI candidate Ta$_2$NiSe$_5$. In the case of exciton-driven order, the amplitude mode of the excitonic condensate leads to very sharp peaks in the 2DCS signal at $(\omega_\tau,\omega_t)=(\pm \omega_{|\Delta|},\omega_{|\Delta|})$ and ($0,\omega_{|\Delta|}$), where $\omega_{|\Delta|}$ corresponds to the largest absorption in the linear optical conductivity.  Tracking the growth of these NR/R/PP signals as a function of temperature, applied pressure, or other tuning parameters can reveal the transition of the material into the ordered phase and the crossover from exciton- to phonon-dominated order. Another signature of the condensate is the massive {\it relative} phase mode peak at $(\omega_\tau,\omega_t)=(-\omega_\psi,\omega_\psi)$.
While our study specialized on Ta$_2$NiSe$_5$, the findings concerning the dominant (hidden) amplitude mode contribution in the 2DCS (conductivity) signal 
and on the selectivity for Coulomb-driven order are generic. They can help to clarify the nature of the insulating phase in other EI candidates~\cite{kaneko2021}, such as Ta$_2$Pd$_3$Te$_5$~\cite{zhang2024,hossain2025,huang2024,yao2024,yi2025} and TiSe$_2$~\cite{cercellier2007,monney2010,kogar2017}. Our results can also be applied to excitonic condensates in artificial heterostructures~\cite{Nandi2012,nguyen2025,qi2025,ma2021}, and more generally to amplitude modes in systems with spontaneous symmetry breaking, including superconductors, charge-density waves and antiferromagnets.

\begin{acknowledgments}
{\it Acknowledgements}
J.C. acknowledges discussions with Y. Shao and Z. Li. J.M. acknowledges interesting discussions with K. Katsumi and S. Giuli. 
J.C. and P.W. were supported by the Swiss National
Science Foundation through the Research Unit QUAST
of Deutsche Foschungsgemeinschaft (FOR5249).   D.G. is supported by the Slovenian Research
and Innovation Agency (ARIS) under Programs No. P1-0044, No. J1-2455, and No. MN-0016-106. J.M. acknowledges support by P1-0044. The calculations were performed on the Beo05 cluster at the University of Fribourg using a code based on~\verb|NESSi|~\cite{nessi}.
\end{acknowledgments}

\bibliography{reference.bib}
\onecolumngrid
\appendix
\newpage

\section{
	\large\large{2D coherent spectroscopy signatures of excitonic condensation in Ta$_2$NiSe$_5$\\
	Supplemental Material}}
\author{Jiyu Chen}
\affiliation{Institute of Physics, Chinese Academy of Sciences, Beijing 100190, China}
\affiliation{Songshan Lake Materials Laboratory, Dongguan, Guangdong 523808, China}
\affiliation{Department of Physics, University of Fribourg, 1700 Fribourg, Switzerland}
\author{Jernej Mravlje}
\affiliation{Jožef Stefan Institute, Jamova 39, SI-1000 Ljubljana, Slovenia}
\affiliation{Faculty of Mathematics and Physics, University of Ljubljana, Jadranska 19, 1000 Ljubljana, Slovenia} 
\author{Denis Gole\v z}
\affiliation{Jožef Stefan Institute, Jamova 39, SI-1000 Ljubljana, Slovenia}
\affiliation{Faculty of Mathematics and Physics, University of Ljubljana, Jadranska 19, 1000 Ljubljana, Slovenia} 
\author{Philipp Werner}
\affiliation{Department of Physics, University of Fribourg, 1700 Fribourg, Switzerland}

\setcounter{figure}{0}
\renewcommand{\thefigure}{S\arabic{figure}}

\section{Model}
For a realistic description of Ta$_2$NiSe$_5$, we use a two-dimensional tight-binding model with six bands and nearest-neighbor hopping amplitudes extracted from density functional theory calculations. The tight-binding Hamiltonian is taken from Ref.~\cite{mazza2020,chatterjee2025}, and the parameters are listed in Tab.~I of the Supplemental Material of Ref.~\cite{chatterjee2025}. In momentum space, the interaction term can be expressed as
$$ \hat H_\text{int}=\sum_{i,j=1}^6\sum_{\textbf{k},\textbf{q}} V_{\textbf{q}}^{ij}  \hat n_{i\textbf{k}} \hat n_{j\textbf{k+q} },$$ where $\hat n_{i\textbf{k}}=\sum_{\textbf{q}\sigma} \hat c_{i \sigma \textbf{k+q}}^{\dagger} \hat c_{i\sigma \textbf{q}}$ is the density operator for orbital $i$ with momentum $\mathbf{k}=(k_x,k_y)$. The interaction matrix for the six-orbital model is
\begin{equation}
	\label{Eq.int} 
V_{\textbf{k}}=\begin{pmatrix}
    U/2 & 0 & 0 &0 &V(k_x) & 0\\
    0 & U/2 & 0 &0 &V(k_x) & 0\\
    0 & 0 & U/2 &0 &0 & V(-k_y)\\
    0 & 0 & 0 &U/2 &0 & V(-k_y)\\
    V(-k_x) & V(-k_x) & 0 &0 &U/2 & 0\\
    0 & 0 & V(k_y) &V(k_y) &0 & U/2\\
\end{pmatrix},
\end{equation}{
with $V(k)=V(1+\exp(\I k))$.

Recent Raman studies~\cite{kim2020,volkov2021,volkov2021failed,ye2021,kim2021} showed that the $B_{2g}$ mode lattice distortion (indicated in Fig.~\ref{fig1}(a) of the main text) dominates the structural phase transition from the orthorhombic to the monoclinic phase.

Here, we model the electron-lattice coupling using a Peierls-like model, where the lattice distortion modifies the overlap between orbitals depending on the relative distance between the atoms. For example, in real space, we can write the couplings between one Ni \#5 atom and two \#1 Ta atoms labeled in Fig.~\ref{fig1}(a) as
\begin{equation}
	g \sum_{\sigma} \hat X (\hat c_{1\sigma }^{\dagger} \hat c_{5\sigma}+\hat c_{1'\sigma  }^{\dagger} \hat c_{5\sigma }),
\end{equation}
where  $g$ is the electron-phonon coupling strength and the operator $\hat {X}$ defines the phonon displacement in the B$_{2g}$ mode. Equation~\eqref{Eq:elph} of the main text takes into account all the orbitals and sites in the lattice model.

\section{Method: Time-dependent Hartree-Fock}
We employ the time-dependent Hartree-Fock method to study the time evolution induced by a laser excitation. The Hartree energy is determined by the local orbital fillings,
 \begin{equation}
 	\label{eq:h}
	\Sigma^H_{ii}(t)=
	U n_{i}(t)+ 2\sum_jV^{ij} n_{j}(t),
\end{equation}
where $U$ is the intra-orbital interaction and $V^{ij}$ is the inter-orbital interaction between orbital $i$ and $j$.
The Fock term is momentum-dependent
\begin{equation}
	\label{eq:f}
  \Sigma_{ij,\mathbf{k}}^F(t)=-\sum_{\mathbf{q}} V^{ij}_{\mathbf{q}} \langle \hat \phi_{ij,\mathbf{k-q}}\rangle(t),
\end{equation}
with $\hat \phi_{ij,\mathbf{k-q}}= \pm \sum_{\sigma}\hat c_{i\sigma\mathbf{k}}^{\dagger} \hat c_{j\sigma\mathbf{q}}(1+\exp(\pm\I (k_x-q_x)))$. For orbital pair $\{ij\}=15,25~(36,46)$,
 the sign is $+$ $(-)$.

We also treat the electron-phonon coupling at the mean-field level,
\begin{equation}
	 	\label{eq:elph}
\hat{H}_{\text{el-ph}}=
-g \langle \hat X\rangle \left[\hat\Delta+ h.c.\right]- g \hat X \left[ \langle\hat\Delta\rangle + h.c.\right],
\end{equation}
where the operator $\hat \Delta$ is given by $\hat\Delta = \hat \phi_{15}-\hat \phi_{25}-\hat \phi_{36}+\hat \phi_{46}$.

The Hartree-Fock equations are first solved self-consistently in equilibrium. For the real-time evolution, we propagate the density matrix as $\rho(t+dt)=e^{-\I \hat{H}^{}_{\text{eff}}[\rho] dt} \rho(t)e^{\I \hat{H}^{}_{\text{eff}}[\rho] dt},$ where $\hat{H}^{}_{\text{eff}}[\rho]$ is the effective Hamiltonian, which includes the kinetic tight-binding Hamiltonian defined in Eq.~\eqref{eq:tb} of the main text, the Hartree-Fock self-energy in Eqs.~\eqref{eq:h}~and~\eqref{eq:f}, and the electron-lattice mean-field contribution in Eq.~\eqref{eq:elph}. This defines a set of non-linear equations which are solved self-consistently at each time step.

Several numerical tricks are employed to improve the efficiency and accuracy: 1) Fourier transformation is used to evaluate the Fock term with $N$ momentum points, leading to a $\mathcal{O}(N\log N)$ complexity, and 2) 5th-order polynomial extrapolation and the 5th-order Adams-Moulton method are used for prediction and correction at each timestep.

\section{Dynamic and static Fock evolution}

Figure~\ref{fig_sm1}(a,d) shows the time evolution of the order parameter $\Delta$ for the realistic parameters for Ta$_2$NiSe$_5$ ($V=0.784$~eV, $g=0.0005$~eV and $\beta=100$~eV$^{-1}$) when we self-consistently evolve the Fock term (Dynamic) or freeze it to the initial value (Static). A single electric pulse is applied, with the same pulse shape as used in the main text.
We notice that both the evolution of $|\Delta|$ and $\theta=\text{arg}(\Delta)$ is dominated by the phase mode. The amplitude mode only makes a small contribution to the $|\Delta|$ signal and is quickly damped, as is best visible in the system with intermediate electron-phonon coupling $g=0.0025$~eV (panel (b)). In the static Fock simulations, the phase mode is suppressed, but the signal shows fast oscillations at an energy identical to that of the amplitude mode. This is because of the sharp feature in the optical conductivity associated with an almost flat band of mixed Ta/Ni character (Fig.~1(c) in the main text). The slowest oscillation visible in the calculations with intermediate and strong electron-phonon coupling (panels (h,i)) corresponds to the phonon frequency $\omega_0=8$~meV. Note that as $g$ increases, the amplitude of the phase mode oscillations is strongly suppressed.

An interesting observation is that for small phonon coupling $g$, the low-energy peak visible in the optical conductivity is actually not coming from the phase mode $\theta$ of the order parameter $\Delta$ defined in the main text, but from a relative phase oscillation $\psi$  of the hopping terms. The operator which is coupled to the $B_{2g}$ phonon in our model is $\hat\Delta=\hat \phi_{15}-\hat\phi_{25}-\hat\phi_{36}+\hat\phi_{46}$. Let us define the phases $\phi_{ij}$ of the expectation values of the operators $\hat\phi_{ij}$, $\langle \hat\phi_{ij} \rangle = \phi_{ij} \equiv |\phi_{ij}|e^{i\theta_{ij}}$ and $\psi \equiv \theta_{15}-\theta_{25} = \theta_{36}-\theta_{46}$. The oscillations of $\psi$, illustrated in Fig.~\ref{fig_sm1}(j,k,l), are analogous to the Leggett mode in multiband superconductors \cite{leggett1966}. There obviously exists a coupling between the relative phase mode (dominant oscillations with $\omega_\psi$) and the amplitude mode (superimposed fast oscillations with $\omega_{|\Delta|}$).

In the exciton-dominated regime (for small electron-phonon coupling) the frequency of the relative phase mode $\omega_\psi$ is higher than the frequency of the massive phase mode $\omega_\theta$ (Fig.~1(e) in the main text), and the low-energy feature visible in the optical conductivity corresponds to the relative phase mode. 
To demonstrate this, we present in Fig.~\ref{fig_sm2} the optical conductivity (panels (a,b)) and the Fourier transforms of the absolute value of the order parameter $|\Delta|$ (panels (c,d)), the phase of the order parameter $\theta$ (panels (e,f)) and the relative phase $\psi$  (panels (g,h)) for $g=0.0005$ and $g=0.0025$. The low-energy peak in the conductivity shows up at $0.04$~eV for $g=0.0005$ and $0.054$~eV for $g=0.0025$ (panels (a,b)).
This energy matches the dominant peak in $|\psi(\omega)|$, while the frequency of the phase mode is more than a factor of two lower. We can also see the strong coupling of the relative phase mode to the amplitude mode in panels (g,h), where the spectra of the dynamic calculation feature a clear peak at the amplitude mode energy.

\begin{figure}[h]
	\includegraphics[width=0.99\linewidth]{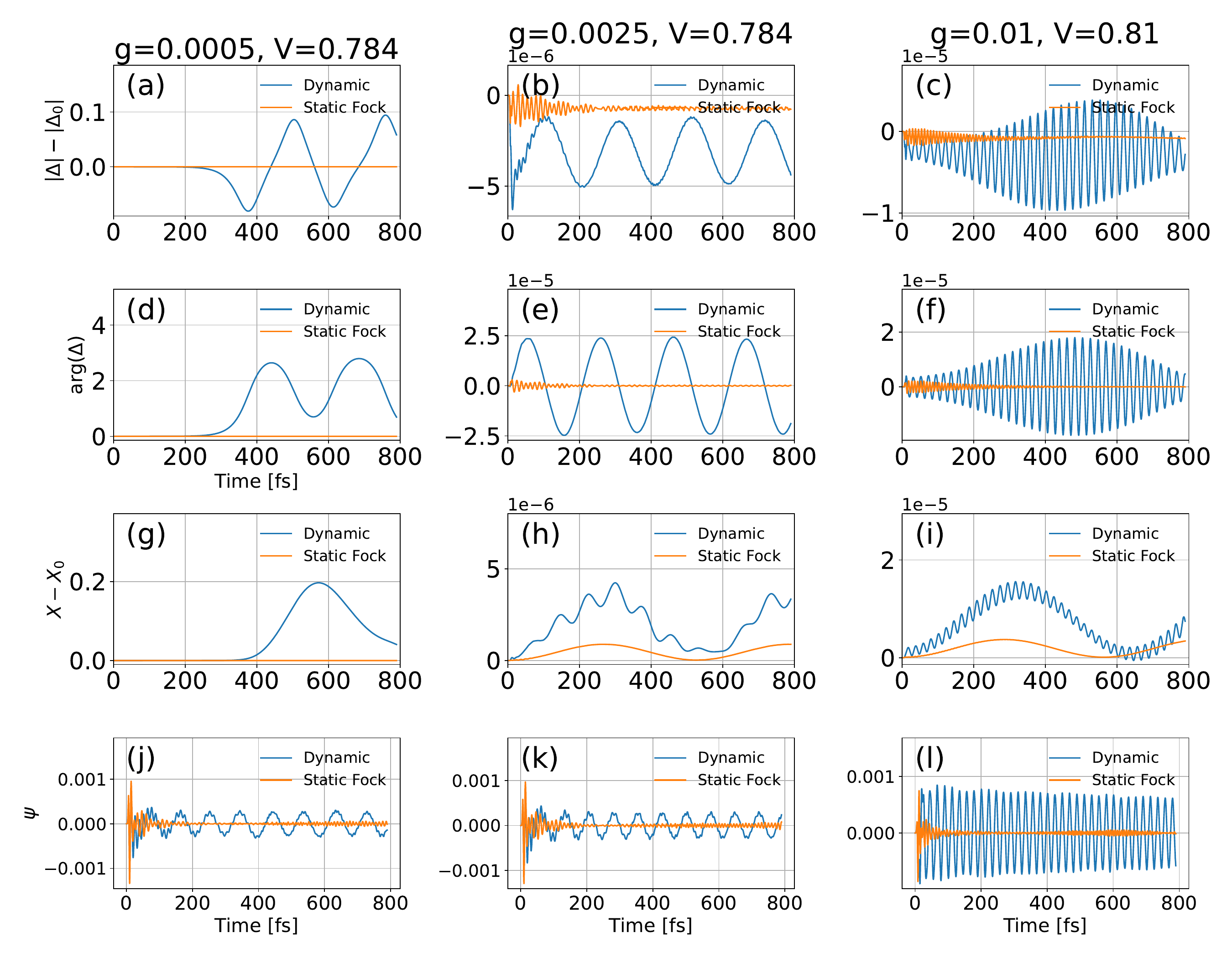}
	\caption{
		Time evolution of the amplitude (a,b,c) and phase (d,e,f) of the order parameter $\Delta$, the phonon displacement $X$ (g,h,i) and the relative phase $\psi$ (j,k,l) obtained with a dynamic and static Fock simulation. The results in panels (a,d,g,j) are for the same parameters as in Fig.~1 of the main text ($V=0.784$~eV, $g=0.0005$~eV and $\beta=100$~eV$^{-1}$).
		The results in panels (b,e,h,k) are for $V=0.784$~eV, $g=0.0025$~eV and $\beta=100$~eV$^{-1}$. 
		The results in panels (c,f,i,l) are for $V=0.81$~eV, $g=0.01$~eV and $\beta=100$~eV$^{-1}$ (same as for Fig.~5(b) in the main text). 
		}
	\label{fig_sm1}
\end{figure}

\clearpage

\begin{figure}[h]
	\includegraphics[width=0.7\linewidth]{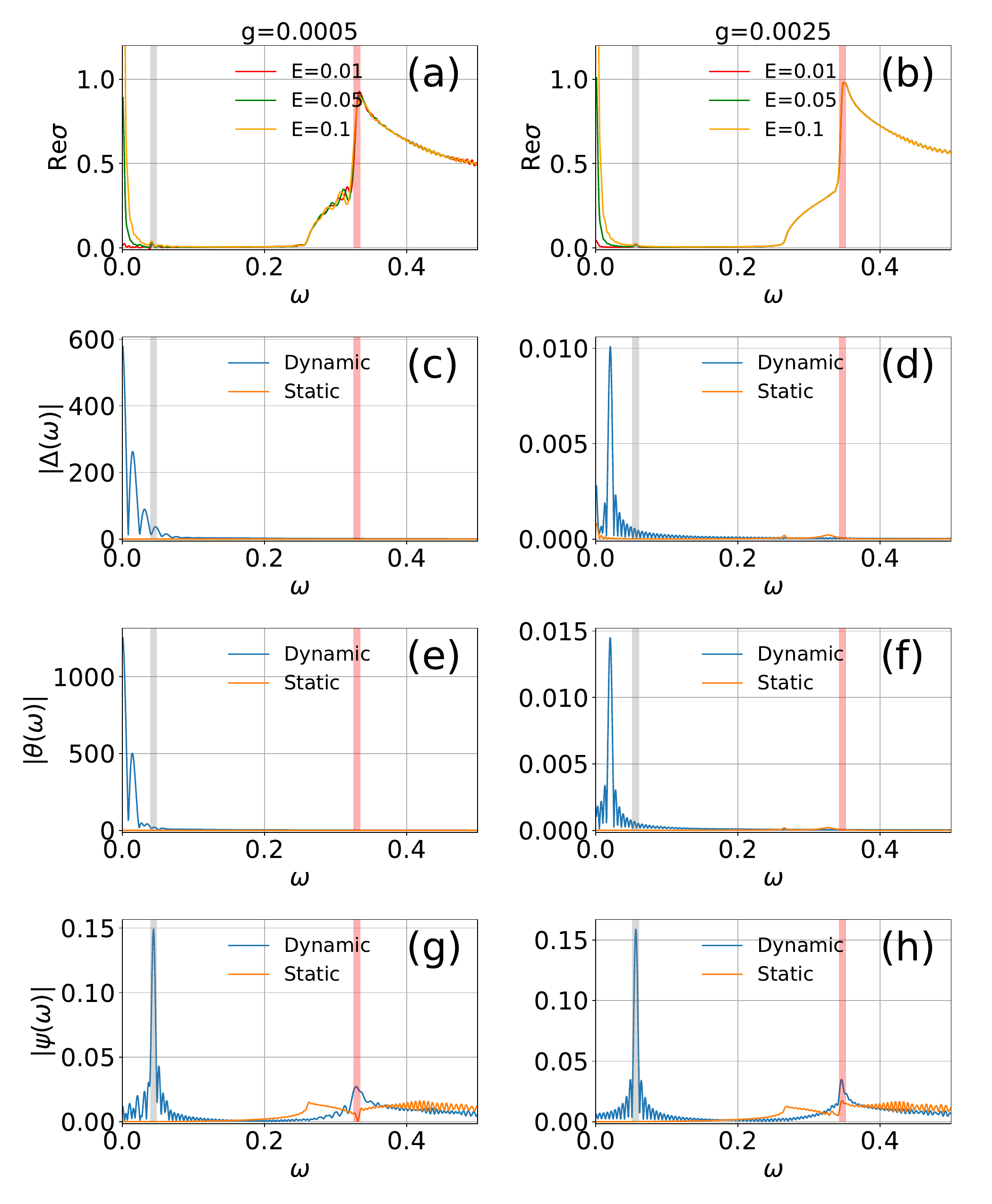}
	\caption{Optical conductivity (a,b) and Fourier transformation of the order parameter $\Delta$ (c,d), phase of the order parameter $\theta$ (e,f) and relative phase $\psi$ (g,h)  with a dynamic and static Fock simulation. The results in panels (a,c,e,g) are for the same parameters as in Fig.~1 of the main text ($V=0.784$~eV, $g=0.0005$~eV and $\beta=100$~eV$^{-1}$). The results in panels (b,d,f,h) are for $V=0.784$~eV, $g=0.0025$~eV and $\beta=100$~eV$^{-1}$. The vertical gray lines indicate the frequency of the relative phase mode $\omega_\psi$ and the red lines the energy of the amplitude mode $\omega_{|\Delta|}$.
	}
	\label{fig_sm2}
\end{figure}

\section{Bandstructure for increasing temperature and electron-phonon coupling}

Figure~\ref{fig_sm3} shows the bandstructures for $g=0.0005$~eV (exciton-dominated order) and $g=0.0025$~eV (phonon-dominated order) for $V=0.784$~eV and different inverse temperatures. The crossover between the two regimes is not directly evident in the bandstructures.

\begin{figure}[t]
\includegraphics[width=0.9\linewidth]{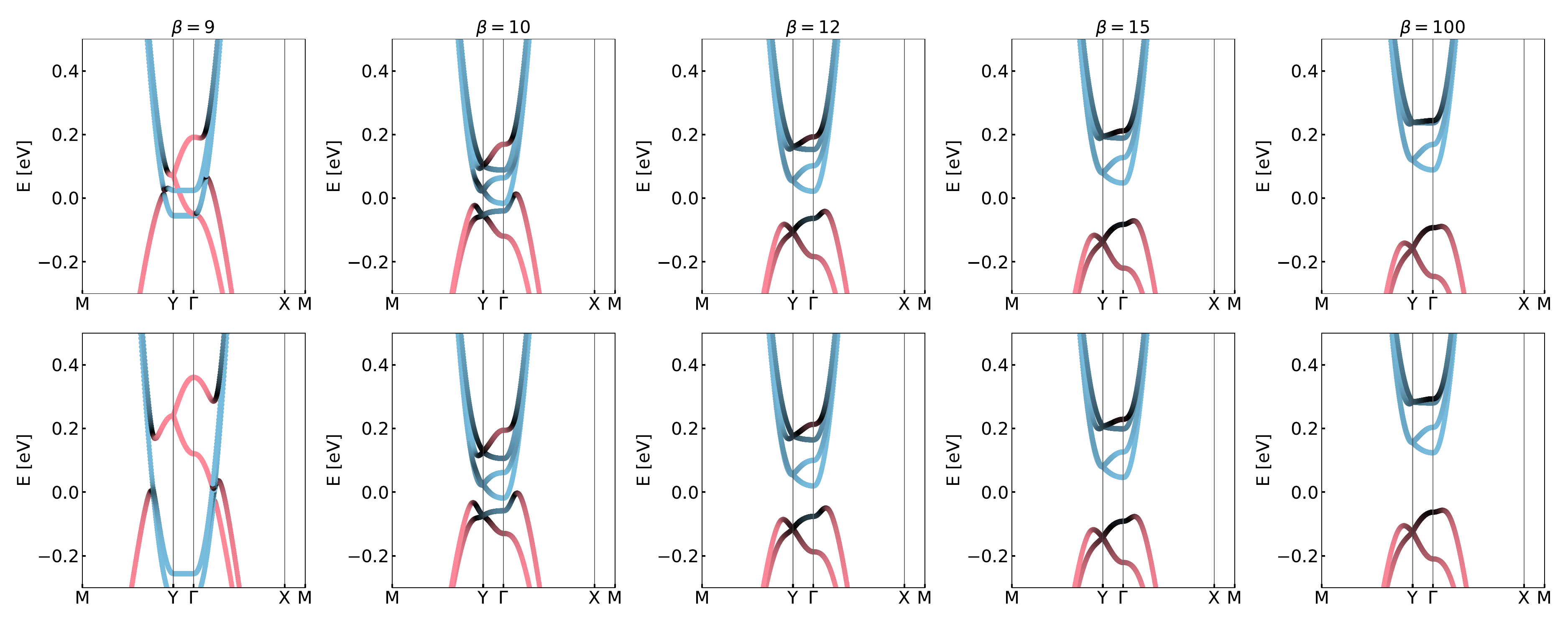}
\caption{Bandstructures for the electron-phonon couplings $g=0.0005$~eV (top) and $g=0.0025$~eV (bottom),  $V=0.784$~eV and the indicated inverse temperatures (in eV$^{-1}$). The blue (red) shading encodes the weight of Ta (Ni) orbitals.
}
\label{fig_sm3}
\end{figure}

\section{Optical conductivity and permittivity}

In Fig.~\ref{fig_sm4}, we present the optical conductivity Re$\sigma(\omega)$ and permittivity Re$\epsilon(\omega)=\text{Re}[1+4\pi i/\sigma(\omega)]$~\cite{larkin2017} for electron-phonon coupling $g=0.0005$~eV and $g=0.0025$~eV with $V=0.784$~eV. The low-temperature optical response shows a shoulder-like feature at the lower edge of the gap. As temperatures increases, the gap size decreases concomitant with the position of the strong resonance, which we associated with the response of the amplitude mode. The in-gap peaks associated with the phase and relative phase modes
also shift to lower energies with increasing temperature, and their weight strongly increases. Similarly, the permittivity shows a divergence at low frequencies, which is enhanced with increasing temperature.

\begin{figure}[t]
	\includegraphics[width=0.8\linewidth]{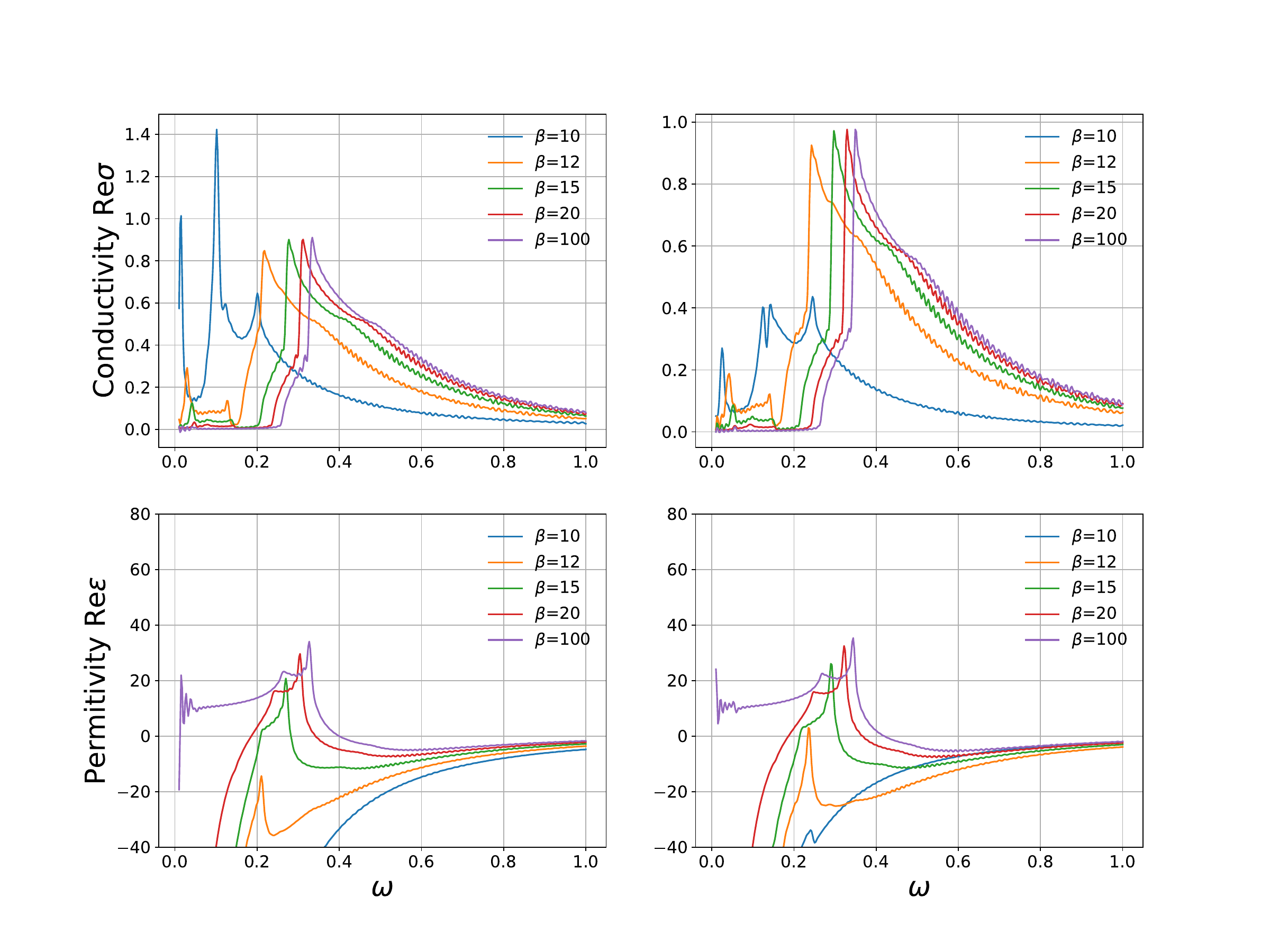}
	\caption{Optical conductivity Re$\sigma$ and permittivity Re$\epsilon$ for electron-phonon coupling $g=0.0005$~eV (left panels) and $g=0.0025$~eV (right panels) with $V=0.784$~eV and the indicated inverse temperatures $\beta$.}
	\label{fig_sm4}
\end{figure}

\end{document}